%% file: main.tex
\documentclass{lanl-report}
\usepackage[utf8]{inputenc}
\usepackage{amsmath}
\usepackage{pdfpages}

\usepackage[usestackEOL]{stackengine}
\usepackage{xcolor, graphicx}

\usepackage{tikz}
\usepackage{changepage}
\usepackage{graphicx}
\usepackage{isotope}
\usepackage{amsmath}
\usepackage{epstopdf}
\usepackage{fancyhdr}
\usepackage{hyperref}
\usepackage{float}
\usepackage{tablefootnote}
\usepackage{stmaryrd}

\usepackage{caption}
\usepackage{subcaption}

\usepackage[justification=centering]{caption}
\RequirePackage{subfiles}
\usepackage[utf8]{inputenc}
\usepackage[backend=biber]{biblatex}
\usepackage{lipsum}

\usepackage{mathtools}
\usepackage{amssymb}
\usepackage{amsthm}
\usepackage{amsmath}
\usepackage{graphicx}

\title{Detector Response Matrices, Effective Areas, and Flash-Effective Areas for Radiation Detectors}
\subtitle{\vspace{0.005\textheight}\\}
\author{
Gregory S. Bowers$^a$\href{https://orcid.org/0000-0002-9524-2234}{\includegraphics[scale=0.5]{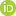}}, 
Eve A. Chase\href{https://orcid.org/0000-0003-1005-0792}{\includegraphics[scale=0.5]{figs/orcid_16x16.png}}, 
William P. Ford\href{https://orcid.org/0000-0001-9946-1226}{\includegraphics[scale=0.5]{figs/orcid_16x16.png}}, 
Daniel D. S. Coupland\href{https://orcid.org/0000-0002-3333-693X}{\includegraphics[scale=0.5]{figs/orcid_16x16.png}}, 
Brian A. Larsen\href{https://orcid.org/0000-0003-4515-0208}{\includegraphics[scale=0.5]{figs/orcid_16x16.png}}, 
Caleb Roecker\href{https://orcid.org/0000-0002-0268-1212}{\includegraphics[scale=0.5]{figs/orcid_16x16.png}}, 
Karl Smith\href{https://orcid.org/0000-0003-2740-5449}{\includegraphics[scale=0.5]{figs/orcid_16x16.png}}, 
Kurtis Bartlett\href{https://orcid.org/0000-0001-6501-4153}{\includegraphics[scale=0.5]{figs/orcid_16x16.png}},
Katherine Gattiker \href{https://orcid.org/0000-0001-8359-1931}{\includegraphics[scale=0.5]{figs/orcid_16x16.png}}, 
Katherine Mesick \href{https://orcid.org/0000-0001-6138-1474}{\includegraphics[scale=0.5]{figs/orcid_16x16.png}}\\ {\footnotesize  $^a$gsbowers@lanl.gov}}
\LAUR{LA-UR-25-31843}

\begin{document}

\includepdf[pages=-]{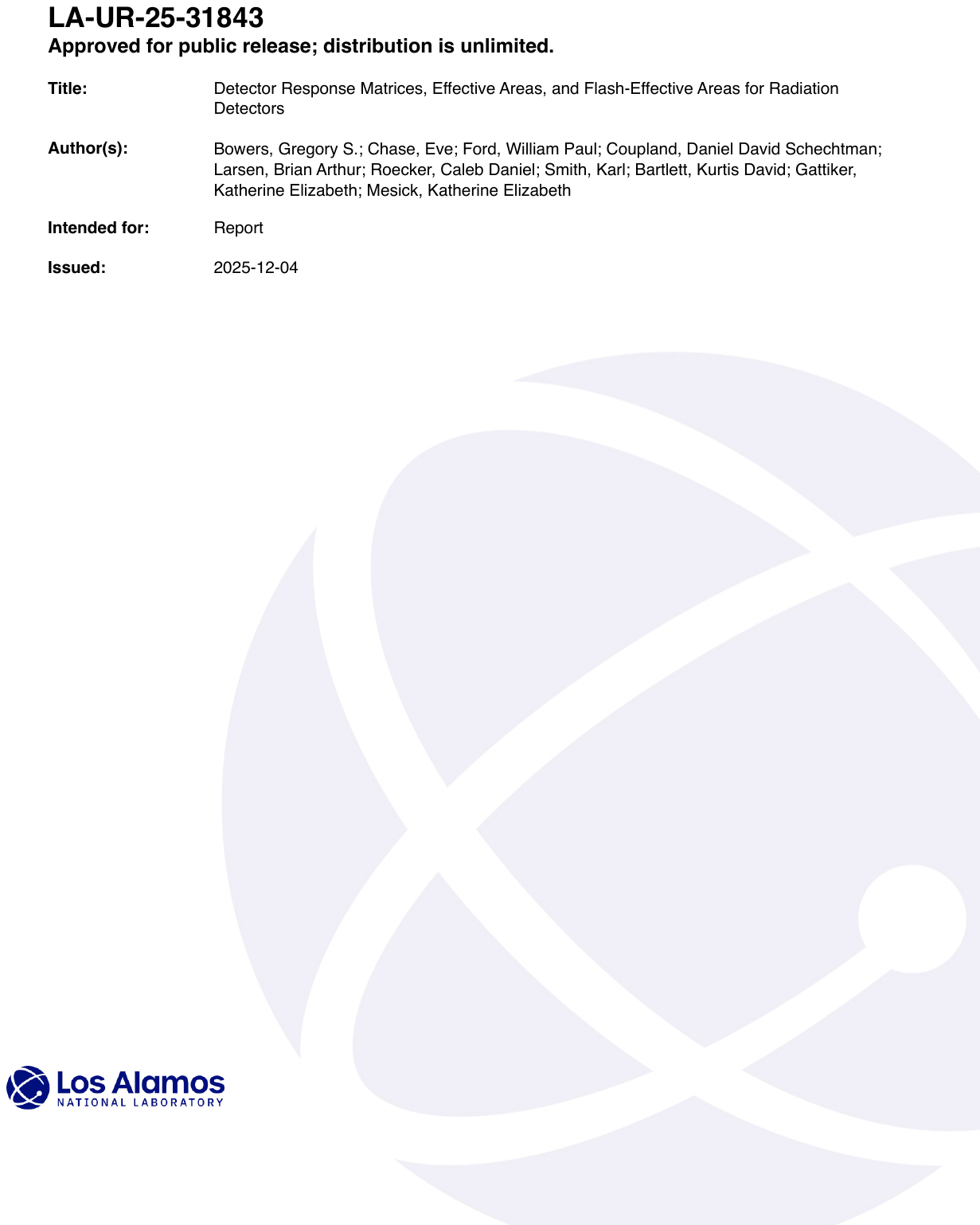}

\LANLTitlePage{}{}

\tableofcontents

\clearpage
\section{Overview}
\subfile{sections/overview}

\section{Introduction}
\subfile{sections/introduction}

\section{Counting Response Function, 
\texorpdfstring{$\mathbf{G}_{\varphi}(E_\mathrm{in}, E_\mathrm{dep})$}{DRF}}
\subfile{sections/counting_response_function}

\section{Counting Response Matrix, 
\texorpdfstring{$\mathbf{G}_{\varphi}^{(j,k)}$}{DRM}}
\subfile{sections/counting_response_matrix}

\section{Counting Effective Area,
\texorpdfstring{$\mathbf{A}_{\varphi}(E_\mathrm{in},[E_A, E_B]_\mathrm{dep})$}{Aeff}}
\subfile{sections/counting_effective_area}

\section{Discrete Effective Area, 
\texorpdfstring{$\mathbf{A}^{(k)}_{\varphi, [E_A, E_B]_\mathrm{dep}}$}{discreteAeff}}
\subfile{sections/discrete_counting_effective_area}

\section{Flash Effective Area, 
\texorpdfstring{$\mathbf{F}_{\varphi}(E_\mathrm{in})$}{FvarphiEin}}
\subfile{sections/flash_effective_area}

\section{Discrete Flash Effective Area,
\texorpdfstring{$\mathbf{F}^{(k)}_\varphi$}{Fkvarphi}}
\subfile{sections/discrete_flash_effective_area}

\section{The Tally Matrix, \texorpdfstring{$n_{j,k}$}{njk}}
\subfile{sections/tally_matrix}

\section{Total Average Effective Area}
\subfile{sections/total_avg_effective_area}

\ShowBibliography

\end{document}

%% file: sections/overview.tex
A Detector Response Matrix (DRM) is a discrete representation of an instrument's Detector Response Function (DRF), which quantifies how many discrete energy depositions occur in a detector volume for a given distribution of particles incident on the detector.  For simple radiation detectors that can count such energy depositions (such as scintillators, Proportional Counter Tubes (PCTs), etc),  we consider the ideal \textbf{counting DRF}, $\mathbf{G}_\varphi (E_\mathrm{in}, E_\mathrm{dep})$, which relates the detector's counting histogram (number of energy depositions within a given channel) to an incident particles  characterization, $\varphi$ (e.g. incident flux, fluence, intensity).  From the counting DRF we can derive the counting DRM, the effective area, and the flash effective area (which measures the total energy deposited in the detector from a large, instantaneous fluence).   

This work is a simplification of the detector efficiency defined in Sullivan 1971 \cite{sullivan1971}; here we assume a fixed incident angular distribution.  While this approach does not consider a fully generalized detector efficiency, $\mathbf{G}_\varphi (E_\mathrm{in}, E_\mathrm{dep}, \Omega)$, and is therefore not suitable to particle telescopes (where one wants to discriminate response based on variable look direction, see \cite{sullivan1971}), it illustrates the fundamental concepts and intuition behind these response functions, their discretization, and provides insight for how to extend to more generalized response functions. 

Finally, this work assumes ideal detectors (able to perfectly resolve energy depositions), and ignores many practical aspects of detector characterization, such as pulse shaping, dead-time, quenching, pulse shape discrimination, etc.  

%% file: sections/introduction.tex
\begin{figure}
\begin{center}
\includegraphics[width=0.75\linewidth, keepaspectratio=true]{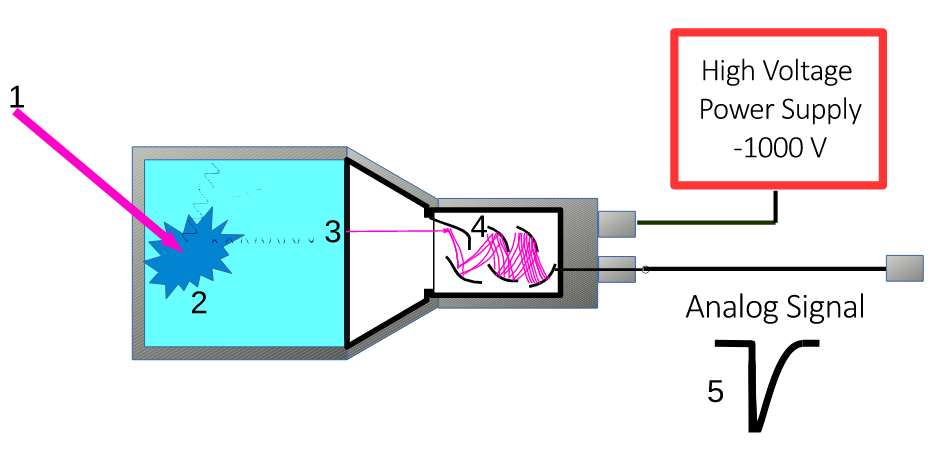}
\caption[Basic scintillation detector geometry.]{\label{fig:Det} Basic scintillation detector.  An incident particle of ionizing radiation characterized by its incident energy, $E_{\mathrm{in}}$ (1), deposits an amount of energy $E_{\mathrm{dep}}$, in a volume of scintillator (cyan) and releases scintillation light (2).   Some of this light travels through the scintillator and causes the emission of a photo-electron (3) at the cathode of an attached photomultiplier tube (PMT).  This photo-electron cascades and multiplies through the high voltage dynode chain (4) and produces an analog current signal at the output (5), which can be input into a multichannel analyser (MCA).  The size of this analog pulse is proportional to the energy deposition ($E_{dep}$) in the scintillator at (2).}
\end{center}
\end{figure}

Particle detectors do not measure particles, they measure energy depositions.  The basic operating principle of a radiation detector that characterizes energy depositions is illustrated in figure \ref{fig:Det} for a scintillation detector. This concept can be applied to other types of detectors that measure, for example, charge instead of scintillation light, such as a neutron proportional counter tube. Ideally a detector is constructed in such a way that it is only sensitive to the particle type to be detected, e.g., a neutron detector being constructed in such a way that only neutrons create energy depositions. However, often other incident particles can create energy depositions in that sensitive volume as well but techniques like pulse-shape discrimination (PSD) may be used to disambiguate them, but again, this is looking at variations in energy depositions.

The most general characterization of $N_\mathrm{in}$ (number of particles in the environment incident on a detector) is given by  the particle radiance spectrum $\dot\Phi_{E,\Omega}$, sometimes referred to particle intensity $J$, according to   

\begin{equation}
\dot\Phi_{E,\Omega} \equiv \dfrac{\mathrm{d}{N_\mathrm{in}}}{\mathrm{d}t\, \mathrm{d}A\, \mathrm{d}E_{in}\mathrm{d}\Omega }~.
\end{equation}

\begin{figure}
\begin{center}
\includegraphics[width=0.60\linewidth, keepaspectratio=true]{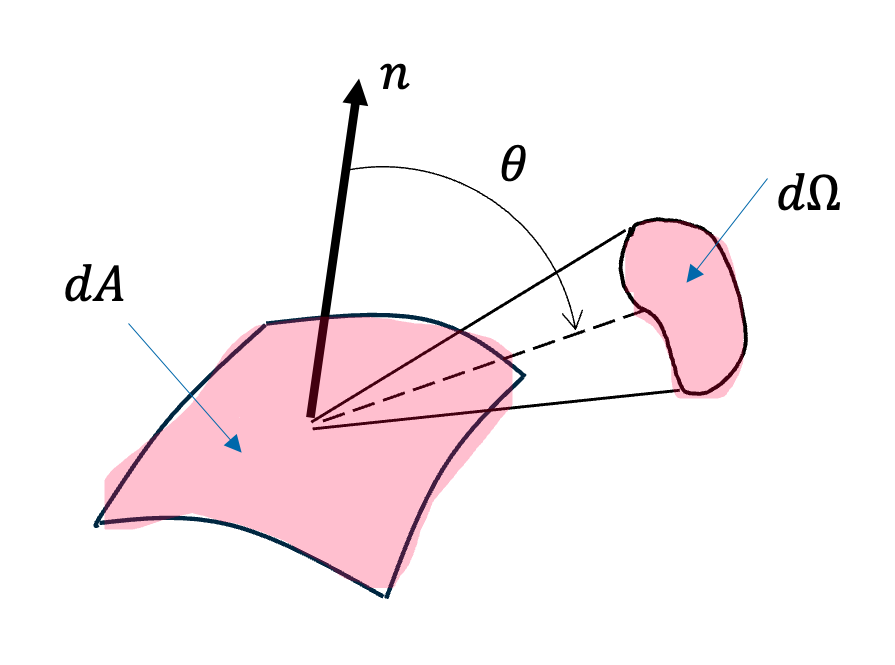}
\caption[Geometry for describing intensity, $J$ of particles at a point in space.]{\label{fig:Flux} Geometry for describing intensity, $J$ of particles at a point in space.  Adapted from ``Radiative Processes in Astrophysics" by Rybicki and Lightman \cite{1979rpa..book.....R}.}
\end{center}
\end{figure}

For example, consider the electron background in the Earth's radiation belts. These electrons will gyrate, bounce, and drift along magnetic field lines.  For a given point in the radiation belts, one can ask the question: how many electrons traverse through a small surface area $\mathrm{d}A$ along a direction $\theta$ relative to the surface normal, confined within a solid angle $\mathrm{d}\Omega$ per unit time $\mathrm{d}t$ per unit particle energy $\mathrm{d}E_\mathrm{in}$, as shown in figure \ref{fig:Flux}.

Different integrations or moments of this relation yield different quantities defined in table \ref{tab:defs}.  For example, for a given angular spectral flux, the number of particles over ranges of time, $t$, incident surface area, $S$, solid angle, $\Omega'$, and incident energy, $E_\mathrm{in}$, are 

\begin{align}
    N_\mathrm{in}  &= \int \mathrm{dN_\mathrm{in}} \\
    &= \int_{t}\int_{S}\int_{\Omega'}\int_{E_{in}=0}^{\infty} \dot\Phi_{E,\Omega} \mathrm{d}t\hat{\mathbf{n}}\cdot\hat{r}\mathrm{d}A\mathrm{d}\Omega\mathrm{d}E~,
\end{align}
where $\hat{\mathbf{n}}\cdot\hat{r} = \cos\theta$, $\hat{n}$ is the direction of the surface area, and $\hat{r}$ is the direction alone the solid angle d$\Omega$.

It is important to distinguish counts of particles in the background environment (i.e., radiation belt electrons, solar energetic particles, etc) from counts of things recorded by a detector (i.e. particles entering a detector volume, energy depositions within that volume, etc).   Typically, particle environments are characterized by the number of incident particles $N_\mathrm{in}$, and their associated incident energies, $E_\mathrm{in}$; counting detectors count the number of energy depositions, $N_\mathrm{dep}$, within a given energy range, or channel, around $E_\mathrm{dep}$.  It is the purpose of the Detector Response Function or Detector Response Matrix to relate  these quantities for a given incident particle characterization, $\varphi$.

\begin{table}[!hbt]
\centering
\caption{\label{tab:defs}Definitions for specifying particle environment, from ICRU Report 85\cite{10.1093/rpd/ncs077}}
\renewcommand{\arraystretch}{2}
\begin{tabular}{c@{} l l}
\hline
$N=$ & & Number of particles in environment\\
$\dot{N}=$ & $\dfrac{\mathrm{d}N}{\mathrm{d}t}$& Particle rate \\ & & number of particles per unit time $(\mathrm{s}^{-1})$\\
$\Phi=$ & $\dfrac{\mathrm{d}N}{\mathrm{d}A}$ & Particle Fluence\\ & & number of particles crossing a unit area $(\mathrm{cm}^{-2})$\\
$\dot{\Phi}=$ & $\dfrac{\mathrm{d}\Phi}{\mathrm{d}t}$ & Particle Fluence Rate (Flux)\\ & & number of particles crossing a unit area per unit time ($\mathrm{cm}^{-2}$~$s^{-1}$)\\
$\Phi_E=$ & $\dfrac{\mathrm{d}\Phi}{\mathrm{d}E}$ & Particle Fluence Spectrum (Spectral Fluence) \\ & & number of particles per incident energy per unit area ($\mathrm{cm}^{-2}$~$\mathrm{keV}^{-1}$)\\
$\dot\Phi_E=$ & $\dfrac{\mathrm{d}\dot{\Phi}}{\mathrm{d}E}$ & Particle Fluence Rate Spectrum (Spectral Flux)\\ & & number of particles per incident energy per unit area per unit time $(\mathrm{cm}^{-2}$~$\mathrm{keV}^{-1}$~$\mathrm{s}^{-1})$\\
$\dot\Phi_{E,\Omega}=$ & $\dfrac{\mathrm{d}\dot{\Phi}_E}{\mathrm{d}\Omega}$ & Particle Radiance Spectrum (Angular Spectral Flux)\\ & &  particles per incident energy per unit area per unit time per unit \\
& & solid angle $(\mathrm{cm}^{-2}$~$\mathrm{keV}^{-1}$~$\mathrm{s}^{-1}$ $\mathrm{Sr}^{-1}$)\\
& & Also called J, ``particle intensity" \cite{1979rpa..book.....R}\\
$\varphi$ & & Incident particle quantification;  \\
& & Generic variable indicating any incident particle quantification, 
\\ & & (e.g. fluence, flux, spectral flux, angular spectral flux, intensity, etc).\\ 
\hline
\end{tabular}
\end{table}
\clearpage

%% file: sections/counting_response_function.tex
A detector response function, $\mathbf{G}_{\varphi}(E_\mathrm{in}, E_\mathrm{dep})$, is a differentiable quantity that relates the number of energy depositions occurring within a detector volume to some quantification of the incident particle environment, $\varphi$; the particle environment can be quantified in many ways, according to either the incident fluence, $\Phi$, fluence rate, $\dot{\Phi}$, fluence spectrum, $\Phi_{E_\mathrm{in}}$, etc (see table \ref{tab:defs}). 

The units of $\mathbf{G}_{\varphi}(E_\mathrm{in}, E_\mathrm{dep})$ will depend on the incident particle quantification $\varphi$.  In practice, the incident particle quantification should be explicitly used and stated when discussing the response;  for example, if a detector response is constructed for the incident particle fluence spectrum, $\Phi_E$, the corresponding response should be written as $\mathbf{G}_{\Phi_E}(E_\mathrm{in}, E_\mathrm{dep})$, and referred to as the `particle fluence spectrum detector response'.   This will make it clear as to both how the detector response has been defined, and how it should be used in practice, as discussed below.   

The detector response function, $\mathbf{G}_{\varphi}(E_\mathrm{in}, E_\mathrm{dep})$, for the number of energy depositions, $N_\mathrm{dep}$, around energy $E_\mathrm{dep}$, within a detector volume, per unit incident particle quantification, $\hat\varphi$, per unit incident energy, $E_\mathrm{in}$, per unit deposited energy, $E_\mathrm{dep}$, is defined as

\begin{align}\label{eq:G}
\mathbf{G}_{\varphi}(E_\mathrm{in}', E_\mathrm{dep}') \equiv \dfrac{1}{\hat\varphi(E_\mathrm{in}')}\left.\dfrac{\partial^2 N_\mathrm{dep}}{\partial E_\mathrm{in}\partial E_\mathrm{dep}}\right|_{E_\mathrm{in}', E_\mathrm{dep}'}~.
\end{align}

The quantity $\hat\varphi$ is the unit value of $\varphi$.  So for example, if the incident particle characterization is flux, $\varphi \equiv \Phi$, and the value of the flux at $E_\mathrm{in} = E_0$ is 50 particle/cm$^2$, then the ratio $\Phi(E_0)/\hat\Phi(E_0)$ = 50.  

Given an incident particle quantification, the total number of energy depositions, $N_\mathrm{dep}$, occurring in the detector volume with energies between $E_A$ and $E_B$ can then be calculated according to the integral expression 

\begin{align}\label{eq:N}
N_\mathrm{dep}([E_A, E_B]_\mathrm{dep}) =\int_{E_\mathrm{A}}^{E_\mathrm{B}}\int_{0}^{\infty} \mathbf{G}_{\varphi}(E_\mathrm{in}, E_\mathrm{dep})\left[\varphi(E_\mathrm{in}) \right]\mathrm{d}E_\mathrm{in}\mathrm{d}E_\mathrm{dep}
\end{align}

where we have used a shorthand notation to indicate the energy deposition channel, $[E_A, E_B]_\mathrm{dep}$, defined by

\begin{align}
[E_A, E_B]_\mathrm{dep} \equiv E_\mathrm{A} < E_\mathrm{dep} < E_\mathrm{B}
\end{align}

We can think of integral equations like \ref{eq:N} as consisting of a \textbf{\textit{kernel operator}} (a function inside an integral, in this case $\mathbf{G}_{\varphi}(E_\mathrm{in}, E_\mathrm{dep})$), acting on, or simply multiplying a \textbf{\textit{kernel argument}} (the thing inside the square brackets, $[\quad ]$, in this case, $\varphi(E_\mathrm{in})$) \footnote{This view will be helpful later when consider the total energy deposition and flash response.}.

Note that equation \ref{eq:N} is a simplification of the definition for the counting rate spectrum found in Sullivan 1971\cite{sullivan1971}, where here we have integrated out the angular dependence.    

%% file: sections/counting_response_matrix.tex
To derive the Detector Response Matrix, we discretize the integral expression for the total number of energy depositions in equation \ref{eq:N}, and look to express it in compact matrix notation.  

Intuitively, a matrix product is a compact way of expressing a discrete integral expression over a single integral or summation \footnote{We recall that the definition of an $m\times n$ matrix $\mathbf{A}$ times an $n\times1$ vector $\mathbf{X}$ is a $m\times 1$ vector $\mathbf{Y} = \mathbf{A}\mathbf{X}$ with elements given by 
\begin{align}
y^{(j)} = \sum_{k=1}^{n} A^{(j,k)}x^{(k)}
\end{align}}. 
We note that equation \ref{eq:N} is a double integral.  We consider an infinitesimal range of energy depositions 
in the channel $[E^{(j)}_\mathrm{dep}, E^{(j)}_\mathrm{dep} + \Delta E^{(j)}_\mathrm{dep}]_\mathrm{dep}$,
and define the energy deposition histogram, $H_\mathrm{dep}(E_\mathrm{dep})$ according to

\begin{align}
H_\mathrm{dep}(E^{(j)}_\mathrm{dep})&\equiv N_\mathrm{dep}([E^{(j)}_\mathrm{dep}, E^{(j)}_\mathrm{dep} + \Delta E^{(j)}_\mathrm{dep}]_\mathrm{dep})\\
&\equiv \int_{E^{(j)}}^{E^{(j)}_\mathrm{dep} + \Delta E^{(j)}_\mathrm{dep}}\int_{0}^{\infty} \mathbf{G}_{\hat\varphi}(E_\mathrm{in}, E_\mathrm{dep})\varphi(E_\mathrm{in}) \mathrm{d}E_\mathrm{in}\mathrm{d}E_\mathrm{dep}\\\label{eq:H}
&\approx \int_{0}^{\infty} \mathbf{G}_{\varphi}(E_\mathrm{in}, E_\mathrm{dep})\varphi(E_\mathrm{in}) \mathrm{d}E_\mathrm{in}\Delta E_\mathrm{dep}^{(j)}
\end{align}

Equation \ref{eq:H} is now a single integral equation, and now amenable to expression as a matrix product.

\begin{figure}
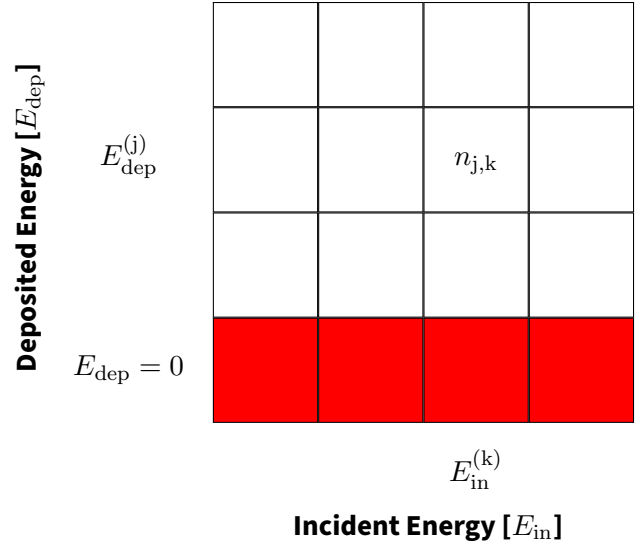

\begin{center}
\raisebox{.5\sqsz}{\rotatebox{90}{\sffamily\bfseries\makebox[-4\sqsz]{Deposited Energy [$E_\mathrm{dep}$]}}}~
\setstackgap{S}{0pt}
\Shortunderstack{ \vsq{}\\ \vsq{$E^\mathrm{(j)}_\mathrm{dep}$}\\ \vsq{} \\ \vsq{$E_\mathrm{dep}=0$}}~~
\stackunder[9pt]{%
  \Shortunderstack{
    \sq{}\sq{}\sq{}\sq{}\\
    \sq{}\sq{}\sq{$n_\mathrm{j,k}$}\sq{}\\
    \sq{}\sq{}\sq{}\sq{}\\
    \sq[red]{}\sq[red]{}\sq[red]{}\sq[red]{}\\
    \hsq{}\hsq{}\hsq{$E^\mathrm{(k)}_\mathrm{in}$}\hsq{}
  }
}{\sffamily\bfseries Incident Energy [$E_\mathrm{in}$]}
\end{center}
\caption[Tally Matrix, $n_\mathrm{j,k}$]{Tally Matrix, $n_\mathrm{j,k}$, of number of energy depositions occurring in detector volume within $E_\mathrm{dep}^{(j)} \le E_\mathrm{dep} < E_\mathrm{dep}^{(j+1)}$, for incident particles with energy within $E_\mathrm{in}^{(k)} \le E_\mathrm{in} < E_\mathrm{in}^{(k+1)}$.  Red bins track number of incident particles that resulted in no energy deposition, $E^\mathrm{(j)}_\mathrm{dep}=0$ (which can also be thought of as `underflow' bins).}\label{fig:tally}
\end{figure}

We define a discrete coordinate grid over the Energy Deposition, $E_\mathrm{dep}$, and Incident Energy, $E_\mathrm{in}$, according to figure \ref{fig:tally}; specifically we use $j$ and $k$ to index into the discrete energy binnings, $E^{(j)}_\mathrm{dep}$, and $E^{(k)}_\mathrm{in}$, respectively. 

We can express the integral form of \ref{eq:H} in discrete form as 

\begin{align} \label{eq:N_discrete}
H_\mathrm{dep}(E^{(j)}_\mathrm{dep}) = \sum_k \mathbf{G}_{\varphi}(E^{(k)}_\mathrm{in}, E_\mathrm{dep}^{(j)}) \varphi(E^{(k)}_\mathrm{in})\Delta E^{(k)}_\mathrm{in}\Delta E^{(j)}_\mathrm{dep}
\end{align}

where the number of discrete energy depositions per energy deposition bin, or counting histogram, is defined as

\begin{align}
\mathbf{H}_\mathrm{dep}^{(j)} = H_\mathrm{dep}(E^{(j)}_\mathrm{dep})
\end{align}

and can be expressed in compact matrix notation according to 

\begin{align}
\mathbf{H}_\mathrm{dep} = \mathbf{G}_\mathbf{\varphi}\boldsymbol{\varphi}
\end{align}

where the detector response matrix, $\mathbf{G}_{\varphi}$ is defined to be 

\begin{align}\label{eq:G_jk}
\mathbf{G}_{\varphi}^{(j,k)} &\equiv \mathbf{G}_{\varphi}(E^{(k)}_\mathrm{in}, E_\mathrm{dep}^{(j)})\Delta E^{(j)}_\mathrm{dep}\Delta E^{(k)}_\mathrm{in}
\end{align}

and the incident particle quantification vector, $\boldsymbol{\varphi}$ is given by

\begin{align}
\boldsymbol{\varphi}^{(k)} &\equiv \varphi(E^{(k)}_\mathrm{in})~.
\end{align}

%% file: sections/counting_effective_area.tex
We can evaluate the $E_\mathrm{dep}$ integral in equation \ref{eq:N} to reduce an integration, and in the process define a new function, or kernel operator, called the effective area, $\mathbf{A}_\varphi$, that is the contribution per incident particle quantification within given energy deposition channel to the total instrument's count rate.

\begin{align}\label{eq:N_Aeff}
N_\mathrm{dep}([E_A, E_B]_\mathrm{dep})
&= \int_{0}^{\infty}\mathbf{A}_\varphi(E_\mathrm{in}, [E_A, E_B]_\mathrm{dep})\varphi(E_\mathrm{in})\mathrm{d}E_\mathrm{in}
\end{align}

where 

\begin{align}\label{eq:Aeff}
\mathbf{A}_\varphi(E_\mathrm{in}, [E_A, E_B]_\mathrm{dep}) \equiv \int_{E_\mathrm{A}}^{E_\mathrm{B}} \mathbf{G}_{\varphi}(E_\mathrm{in}, E_\mathrm{dep}) \mathrm{d}E_\mathrm{dep}~.
\end{align}

By reducing equation \ref{eq:N_Aeff} to a single integration, we can discretize it and reduce it's computation to a compact vector dot product calculation. 

%% file: sections/discrete_counting_effective_area.tex
Discretizing equation \ref{eq:N_Aeff} and \ref{eq:Aeff}, and using the definition of the DRM in equation \ref{eq:G_jk}, we can express the integral counts, as 

\begin{align}
N_\mathrm{dep}([E_A, E_B]_\mathrm{dep}) &= \sum_k \left(\EdepSum \mathbf{G}_{\varphi}(E^{(k)}_\mathrm{in}, E_\mathrm{dep}^{(j)}) \Delta E^{(j)}_\mathrm{dep}\Delta E^{(k)}_\mathrm{in}\right)\varphi(E^{(k)}_\mathrm{in}) \\
&= \sum_k \left(\EdepSum \mathbf{G}_{\varphi}^{(j,k)}\right)\varphi(E^{(k)}_\mathrm{in}) \\
&= \sum_k \mathbf{A}_{\varphi, [E_A, E_B]_\mathrm{dep}}^{(k)} \varphi(E^{(k)}_\mathrm{in})
\end{align}

or recalling the definition of the vector dot product \footnote{if we consider $\mathbf{A}_\varphi$ and $\boldsymbol{\varphi}$ to be column vectors, we can equivalently write the number of energy depositions as $N_\mathrm{dep}( [E_A, E_B]_\mathrm{dep}) = \mathbf{A}_\varphi^{\top} \boldsymbol{\varphi}$.}, 

\begin{align}
N_\mathrm{dep}( [E_A, E_B]_\mathrm{dep}) = \mathbf{A}_{\varphi, [E_A, E_B]_\mathrm{dep}}^{(k)} \cdot \boldsymbol{\varphi}
\end{align}

where 

\begin{align}
\mathbf{A}_{\varphi, [E_A, E_B]_\mathrm{dep}}^{(k)} 
&\equiv \sum_{\substack{j \in \\ E_A < E_\mathrm{dep}<{E_B}}} \mathbf{G}_{\varphi}^{(j,k)}~.
\end{align}

%% file: sections/flash_effective_area.tex
In this section we will develop the concept of a flash effective area, which measures the total energy deposition in our instrument, and is functionally analogous to the counting effective area developed in the previous section.

We start by writing the integral representation for the total energy deposited, $T_\mathrm{dep}$, for an incident particle characterization, $\varphi(E_\mathrm{in})$, given by 

\begin{align}\label{eq:Tdep}
T_\mathrm{dep} = \int_{0}^{\infty}\int_{0}^{\infty} \mathbf{G}_{\varphi}(E_\mathrm{in}, E_\mathrm{dep})\left[\varphi(E_\mathrm{in}) E_\mathrm{dep}\right]\mathrm{d}E_\mathrm{in}\mathrm{d}E_\mathrm{dep}~.
\end{align}

Previously, we defined our effective area as the integral over $E_\mathrm{dep}$ within the expression for the total number of counts (see equation \ref{eq:Aeff}); this produced a kernel operator that could be convolved with the incident particle characterization, $\varphi(E_\mathrm{in})$ (the kernel argument of \ref{eq:N_Aeff}).  For the flash response, we wish to create a kernel operator that can be convolved with the incident energy-characterization, $\varphi(E_\mathrm{in}) E_\mathrm{in} \equiv \varphi(E_\mathrm{in}) \times E_\mathrm{in} $.  We see in equation \ref{eq:Tdep} that we don't quite have the kernel argument we want;  we have $\varphi(E_\mathrm{in}) E_\mathrm{dep}\equiv \varphi(E_\mathrm{in})\times E_\mathrm{dep}$, which is not a well defined physical quantity.  We can get to an incident-energy characterization as the kernel argument of \ref{eq:Tdep} if we introduce the factor $E_\mathrm{in}/E_\mathrm{in} = 1$ within the integral, and group the factor $E_\mathrm{dep}/E_\mathrm{in}$ with the detector response function.  

We therefore define the  the total energy deposition with the following functional form

\begin{align}\label{eq:Tdep_F}
T_\mathrm{dep} = \int_{0}^{\infty} \mathbf{F}_{\varphi}(E_\mathrm{in})\left[\varphi(E_\mathrm{in}) E_\mathrm{in}\right]\mathrm{d}E_\mathrm{in}
\end{align}

then comparing equations\ref{eq:Tdep_F} with \ref{eq:Tdep}, it follows that 

\begin{align}
\mathbf{F}_{\varphi}(E_\mathrm{in}) &\equiv \int_{0}^\infty \mathbf{G}_{\varphi}(E_\mathrm{in}, E_\mathrm{dep}) \left(\dfrac{E_\mathrm{dep}}{E_\mathrm{in}}\right) \mathrm{d}E_\mathrm{dep}~.
\end{align}


By this definition, the flash effective area will have the same units as the counting effective area, and we have created a kernel operator which can be convolved with a physically well defined quantity, the incident-energy characterization, $\varphi(E_\mathrm{in}) E_\mathrm{in}$.

%% file: sections/discrete_flash_effective_area.tex
Discretizing equation \ref{eq:Tdep_F}, we have that the 

\begin{align}\label{eq:Tedep_exact}
T_{\mathrm{dep}} = \sum_k  \mathbf{F}_{\varphi}(E_\mathrm{in}^{(k)}) \; \varphi^{(k)} E^{(k)}_{\mathrm{in}}\Delta E^{(k)}_\mathrm{in}
\end{align}

which can be expressed in compact matrix notation (using the dot product) according to 

\begin{align}
T_\mathrm{dep} = \mathbf{F}_\varphi \cdot (\varphi E_\mathrm{in})
\end{align}

where 

\begin{align}
\mathbf{F}^{(k)}_\varphi \equiv   \mathbf{F}_{\varphi}(E_\mathrm{in}^{(k)})\Delta E^{(k)}_\mathrm{in}
\end{align}

and 

\begin{align}
(\varphi E_\mathrm{in})^{(k)} \equiv \varphi(E^{(k)}_\mathrm{in})E^{(k)}_\mathrm{in}~.
\end{align}

A detector flash response, $\mathbf{F}_\varphi(E_\mathrm{in})$ estimates the average fractional energy deposition per unit incident particle characterization and is computed according to 

\begin{equation}\label{eq:flash}
    \mathbf{F}_\varphi^{(k)} \equiv \dfrac{1}{\varphi^{(k)}}\sum_{i=1}^{N_\mathrm{in}^{(k)}} \left(\dfrac{\mathcal{E}_\mathrm{i,dep}^{(k)}}{E_\mathrm{in}^{(k)}}\right)
\end{equation}

where the summation is over the discrete Monte Carlo events, indicated by the index i, where $\mathcal{E}_\mathrm{i,dep}^{(k)}$ is the energy deposition in the detector resulting from the $i^{\mathrm{th}}$ Monte Carlo throw of incident particles with energy in bin $E_\mathrm{in}^{(k)}$. 

The Flash response in \ref{eq:flash} can be computed exactly using a counting DRM, $\mathbf{G}_{j,k}$ according to 

\begin{align}
    \mathbf{F}_\varphi^{(k)} &\equiv \dfrac{1}{\varphi^{(k)}}\sum_{j} \left(\dfrac{n_{j,k}\langle E_\mathrm{dep}^{(j)}\rangle}{E_\mathrm{in}^{(k)}}\right)\\
    &=\sum_{j}\mathbf{G}^{(j,k)} \left(\dfrac{\langle E_\mathrm{dep}^{(j)}\rangle}{E_\mathrm{in}^{(k)}}\right)
\end{align}

where $\langle E_\mathrm{dep}^{(j)}\rangle$ is the mean energy deposition within bin j.  We can then estimate the total energy deposition in \ref{eq:Tedep_exact} according to 

\begin{align}\label{eq:eq:Tedep_approx}
T_{\mathrm{dep}} = \sum_j\sum_k \mathbf{G}^{(j,k)}_\varphi \;\varphi^{(k)} \langle E_\mathrm{dep}^{(j)}\rangle~.
\end{align}

If we assume that $\langle E_\mathrm{dep}^{(j)}\rangle \approx \frac{1}{2}(E_\mathrm{dep}^{(j)} + E_\mathrm{dep}^{(j+1)})$, i.e. that the average Energy deposted in bin j is the average energy of the bin, j, then the todal energy deposition is approximated by

\begin{align}\label{eq:eq:Tedep_approx}
T_{\mathrm{dep}} \approx \sum_j\sum_k \mathbf{G}^{(j,k)}_\varphi\; \varphi^{(k)}  \dfrac{1}{2}(E_\mathrm{dep}^{(j)} + E_\mathrm{dep}^{(j+1)})~.
\end{align}

%% file: sections/tally_matrix.tex
We define a tally matrix on a grid of incident and deposited energies according to figure \ref{fig:tally} where $j$ is an index into a discrete array of energy deposition bin edges, $E_\mathrm{dep}^{(j)}$; $k$ is an index into a discrete array of incident particle energy bin edges, $E_\mathrm{in}^{(k)}$;  the bin widths $\Delta E^{(i)} \equiv E^{(i+1)} - E^{(i)}$; and $\varphi(E_\mathrm{in}^{(k)})$ is the incident particle characterization within the interval $E_\mathrm{in}^{(k)} < E_\mathrm{(in)} < E_\mathrm{in}^{(k)} + \Delta E_\mathrm{in}^{(k)}$.

The detector response matrix, $\mathbf{G}_\varphi^{(j,k)}$, follows from substitution of the definition of $\mathbf{G}_\varphi^{(j,k)}$ given by equation \ref{eq:G_jk} into the discretized form of equation \ref{eq:G}

\begin{equation}
    \mathbf{G}_\varphi^{(j,k)} = \dfrac{n_{j,k}}{\hat\varphi(E^{(k)}_\mathrm{in})}
\end{equation}

where

\begin{align*}
\varphi(E_\mathrm{in}^{(k)}) &\equiv \dfrac{1}{\Delta E^{(k)}}\int^{E_\mathrm{in}^{(k+1)}}_{E_\mathrm{in}^{(k)}}\varphi(E)\mathrm{d}E\\
\end{align*}

and $n_{j,k}$ is a tally matrix defined as

\begin{align*}
n_{j,k} \equiv &\text{Total number of Energy depositions}\\
&\text{with $E_\mathrm{dep}^{(j)} \le E_\mathrm{dep} < E_\mathrm{dep}^{(j+1)}$ from}\\ 
&\text{incident particles with energies}\\
&\text{with $E_\mathrm{in}^{(k)} \le E_\mathrm{in} < E_\mathrm{in}^{(k+1)}$}\\
\equiv &\sum_{i=1}^{N_\mathrm{in}^{(k)}} \llbracket E_\mathrm{dep}^{(j)} \le E_\mathrm{dep}^{(i)} < E_\mathrm{dep}^{(j+1)}\rrbracket
\end{align*}

where $\llbracket P \rrbracket$ is the Iverson bracket, evaluating to 1 if P is true, and 0 otherwise. 
 
$\mathbf{G}_\varphi^{(j,k)}$ is the number of energy depositions within the range $[E_\mathrm{dep}^{(j)}, E_\mathrm{dep}^{(j+1)}]$ per unit $\varphi$ of incident particles with energies in the range of $[E_\mathrm{in}^{(k)}, E_\mathrm{in}^{(k+1)}]$, per unit energy deposition. 

\subsection{Normalizing the Tally Matrix}

To finish the calculation of the detector response matrix, one needs to \textbf{normalize} the tally matrix $n_\mathrm{j,k}$ to the simulations specific input geometry, and the number of particles thrown, which amounts to calculating the quantities $\varphi(E_\mathrm{in}^{(j)}, E_\mathrm{in}^{(j+1)})$. 

Except for point sources, the simulated quantification of interest for the incident particle environment $\varphi$ will typically depend on the surface from which the incident particles are thrown at the detector (i.e. when $\varphi \sim dN/dA)$  

In general, the detector response matrix will look like 

\begin{equation}
    \mathbf{G}^{(j,k)}_\varphi \sim \dfrac{n_\mathrm{j,k}}{N_\mathrm{j}} A
\end{equation}

where $A$ is an effective input area dependent on the particle input geometry of the simulation.  

We consider two typical cases: 1.) a incident plane wave, 2.) an isotropic (in solid angle) particle field. 

\subsubsection{Simulated Plane Wave}

If particles are thrown from a planer surface with area $A^*$, in the direction of $\theta_\mathrm{in}$ with respect to the planar surface normal, then the effective input area is 

\begin{equation}
    A_\mathrm{in} = A^* \cos\theta_{\mathrm{in}}
\end{equation}

If the number of particles thrown with energies in the range of $E_j < E_\mathrm{in} < E_{j+1}$ is $N_j$, the simulated fluence in this energy range is

\begin{equation}
    \Phi(E_\mathrm{in}^{(j)}, E_\mathrm{in}^{(j+1)}) = \dfrac{N_\mathrm{j}}{A_\mathrm{in}}
\end{equation}

Then for a planewave, the detector response matrix $\mathbf{G}_\Phi^{(j,k)}$ is 

\begin{equation}
    \boxed{
    \mathbf{G}_\Phi^{(j,k)} = \dfrac{n_\mathrm{j,k}}{N_\mathrm{j}} A_\mathrm{in}
    }
\end{equation}

\subsubsection{Simulated Isotropic Flux}

If the detector response to a directionally isotropic particle field is desired, then one typically throws particles from the surface of a sphere of radius $R$ surrounding the detector according to a cosine-law distribution in the solid angle around the input surface normal and constructs the tally matrix $n_\mathrm{j,k}$ as described previously. 

First, we assume that the particle intensity separates into independent energy
and angular components given by

\begin{equation}\label{eq:J}
    \Phi_{E,\Omega}(E,\Omega) = \Phi_{E}(E)F(\Omega)
\end{equation}

A directionally isotropic source implies that $F(\Omega) = 1/4\pi$.

To determine the proper normalization, we calculate the number of incident particles, $N$, input on a sphere of radius $R$, for a particle intensity $\Phi_{E,\Omega}(E,\Omega)=\Phi_{E}(E)F(\Omega)$ with $F(\Omega) = 1/4\pi$.   At the surface the particle intensity in the direction of $\theta$ relative to the surface normal is reduced by the factor $\mathrm{d}\vec{\sigma}\cdot\hat{r}=\cos\theta\mathrm{d}\sigma$.  Then 

\begin{align} \label{eq:norm}
N &=  \int_S \int_{\Omega'} \int_{E=0}^\infty \Phi_{E}(E)F(\Omega) \mathrm{d}E\, \cos\theta \mathrm{d}\Omega\mathrm{d}\sigma \\
  &= \left[\int_S \mathrm{d}\sigma\right] \left[\int_{\Omega'}\cos\theta \mathrm{d}\Omega\right] \left[\dfrac{1}{4\pi}\int_{E=0}^\infty \Phi_E\left(E\right)\mathrm{d}E_\mathrm{in} \right]\\
  &= [4\pi R^2][\pi]\left[\dfrac{\Phi}{4\pi}\right]  \\
  &= 4\pi^2 R^2\mathbf{j}_0  \\
  &= \pi R^2 \Phi 
\end{align}

where $\mathbf{j_0} \equiv \Phi/4\pi$ is sometimes referred to as the ``directional flux", with $\Phi$ refered to as the ``omni-directional flux" \cite{yando2011monte}. 

The omnidirectional flux for particles thrown with energies in the range of $E_j < E_\mathrm{in} < E_{j+1}$ is given by 

\begin{equation}
    \Phi(E_\mathrm{in}^{(j)}, E_\mathrm{in}^{(j+1)}) = \dfrac{N_\mathrm{j}}{\pi R^2}
\end{equation}

and the detector response matrix for an isotropic or omni-directional flux is given by 

\begin{equation}\label{eq:Giso}
    \boxed{
    \mathbf{G}^{(j,k)}_\Phi = \dfrac{n_\mathrm{j,k}}{N_\mathrm{j}} \pi R^2
    }
\end{equation}

We emphasize that equation \ref{eq:Giso} as defined is the detector response matrix that relates the omnidirectional flux, $\Phi$, (counts/cm$^2$/s) to instrument count rate.  If one wanted to relate the instrument count rate to a directional flux $\mathbf{j}_0$, the geometric factor would be given by equation \ref{eq:Giso} times an additional factor of $4\pi$, or $\mathbf{G}^{(j,k)}_{\mathbf{j}_0} = 4\pi \mathbf{G}^{(j,k)}_\Phi$.

%% file: sections/total_avg_effective_area.tex
The \textbf{Total effective area}, $\mathbf{G}$, is the average effective area [cm$^2$] for any energy deposition ($E_k > 0$) over all incident particle energies ($E_j > 0$), and is given by 



\begin{align}\label{eq:GG}
    \langle\mathbf{G}_\varphi\rangle &= \dfrac{1}{(E_{K} - E_{k=0})} \sum\limits_{j,k} \Delta{E}_k \mathbf{G}_\varphi^{\mathrm{(j,k)}}
\end{align}

where $K$ is the total number of incident energy bins, and, $E_K - E_{k=0}$ is the span of incident energies.  
It is also useful to define the total effective area for energy depositions above some threshold energy as $\mathbf{G}(E_{\mathrm{dep}} > E_{\mathrm{thres}})$, where k in equation \ref{eq:G} satisfies $E_k > E_{\mathrm{thres}}$.